\documentclass[pmlr]{jmlr}

\usepackage{longtable}

\usepackage{booktabs}
\usepackage{siunitx}

\usepackage{xcolor}

\usepackage{microtype}

\makeatletter
\def\ps@jmlrheadings{\ps@plain}
\makeatother
\begin{document}

\title[Rapid Response for LLM Protection]{A Framework for Rapidly Developing and Deploying Protection Against Large Language Model Attacks}


\author{\Name{Adam Swanda} \Email{aswanda@cisco.com}\\  
 \Name{Amy Chang} \Email{changamy@cisco.com}\\
 \Name{Alexander Chen} \Email{alexc92@cisco.com}\\
 \Name{Fraser Burch} \Email{burchy@cisco.com}\\
 \Name{Paul Kassianik} \Email{paulkass@cisco.com}\\
 \Name{Konstantin Berlin} \Email{berlink@cisco.com}\\
 \addr Cisco Systems}

\maketitle
\thispagestyle{plain}

\begin{abstract}
The widespread adoption of Large Language Models (LLMs) has revolutionized AI deployment, enabling autonomous and semi-autonomous applications across industries through intuitive language interfaces and continuous improvements in model development.
However, the attendant increase in autonomy and expansion of access permissions among AI applications also make these systems compelling targets for malicious attacks.
Their inherent susceptibility to security flaws necessitates robust defenses, yet no known approaches can prevent zero-day or novel attacks against LLMs. This places AI protection systems in a category similar to established malware protection systems: rather than providing guaranteed immunity, they minimize risk through enhanced observability, multi-layered defense, and rapid threat response, supported by a threat intelligence function designed specifically for AI-related threats.

Prior work on LLM protection has largely evaluated individual detection models rather than end-to-end systems designed for continuous, rapid adaptation to a changing threat landscape.
To address this gap, we present a production-grade defense system rooted in established malware detection and threat intelligence practices.
Our platform integrates three components: a threat intelligence system that turns emerging threats into protections; a data platform that aggregates and enriches information while providing observability, monitoring, and ML operations; and a release platform enabling safe, rapid detection updates without disrupting customer workflows.
Together, these components deliver layered protection against evolving LLM threats while generating training data for continuous model improvement and deploying updates without interrupting production.
We share these design patterns and practices to surface the often under-documented, practical aspects of LLM security and accelerate progress on operations-focused tooling.
\end{abstract}

\begin{keywords}
Large Language Models, LLM Security, Threat Intelligence, Machine Learning Operations, Rapid Response, Guardrails, Adversarial Attacks, Detection Systems
\end{keywords}

\section{Introduction}

Building resilient and adaptive security detection platforms for AI is distinct from traditional detection platforms primarily due to the adaptive nature of attacks and the nondeterministic nature of generative AI applications.
Security detection systems are locked in a continuous cybersecurity OODA loop (observe, orient, decide, act)~\cite{sager_ooda_loop} with their attackers.
The constant engagement means the focus of a detection platform should shift from expending significant one-off effort to achieve the "best" detection model at a particular point in time, and instead focus on rapid adaptation and response.
As soon as a new detection engine is released, its owner must quickly address any new bypasses that are discovered.

Defensive strategies for LLM deployments have consisted of a mix of technical controls and operational best practices~\cite{owasp2025} that span a range of effectiveness and cost-efficiency.
We use ``guardrails'' to refer to the deployed detection and blocking layer that mediates model inputs/outputs and enforces safety and security policies.
Examples include input and output sanitization (e.g., guardrails that filter out malicious prompts and dangerous commands)~\cite{kumar2024certifyingllmsafetyadversarial,jiang2023identifyingmitigatingvulnerabilitiesllmintegrated}, continuous monitoring and logging to detect anomalies and signs of attack~\cite{Christiano2022}, red teaming and stress testing to discover vulnerabilities and observe model behavior~\cite{bullwinkel2025lessonsredteaming100}, and adversarial training to build out inherent defenses against unauthorized manipulation~\cite{jain2023baselinedefensesadversarialattacks}.

LLMs can process vast, unstructured datasets from diverse sources to identify anomalies and patterns.
Machine learning-based detection models tend to generalize better than signatures for detections and form the basis of most modern security detection products.
They are especially important for protecting LLMs from prompt injection attacks~\cite{liu2024formalizingbenchmarkingpromptinjection}—when an adversary prompts a model to override the developers' guardrails and executes a malicious instruction.
Here, semantic understanding of language is critical and pattern-matching approaches alone are not able to fully capture all the nuanced syntactical variations~\cite{greshake2023youvesignedforcompromising}.

However, models often generalize poorly to out-of-distribution (OOD) inputs, including novel attack patterns or techniques against LLMs~\cite{benoit2024unravelingkeycomponentsood}~\cite{peng2024rapidresponsemitigatingllm}.
As a result, attacks not present in the training distribution (e.g., zero-days) are frequently misclassified.
Even when ML models do generalize, their detection accuracy is hampered by operational requirements to deploy these models at very low false positive rates, and further complicated by the low base rate of attacks within legitimate network traffic and user activity~\cite{PIETRASZEK2005169}.
To our knowledge, all current detection systems can be bypassed under some conditions, and there is no known method to prevent adversarial attacks on ML models with 100\% certainty~\cite{peng2024rapidresponsemitigatingllm,RiosInsua03072023}.

Furthermore, it is hard to guarantee ML model consistency across multiple retraining cycles~\cite{wang2020wisdomensembleimprovingconsistency}.
A new model deployment could introduce unexpected changes in behavior, such as blocking a previously unflagged input and disrupting customer workflows.
Therefore, safely releasing an ML model requires careful observation and gradual customer rollout to gather a statistically significant amount of observations.
This validation process can conflict with standard engineering release cycles, which require a much faster quality assurance process.

Given these realities, the only practical solution is ensuring rapid response and timely protection against novel threats through continuous detection updates.
This operational tempo allows security teams to block attacks before they become commoditized and have the potential to compromise a large fraction of users and customers.
A robust platform must be able to rapidly identify new threats, deploy initial basic protections against them, while also enabling longer-term training data for more robust ML detection models.

To address these challenges, we have developed and operationalized a dedicated threat intelligence function that considers expanded attack surfaces and deepening integrations of generative AI and agentic AI tools in software systems and technology.
We have also developed a novel platform architecture that advances the state-of-the-art in LLM security operations through systematic integration of threat intelligence, data-driven decision making, and safe deployment methodologies.
A critical differentiator of our platform is its ability to systematically improve detection capabilities through operational feedback loops that transform real-world threats into enhanced security postures.
Our key research contributions include:
\begin{itemize}
  \item A \textbf{systematic threat intelligence operations capability} that introduces automated prioritization algorithms and novel attack-to-signature translation mechanisms for rapid zero-day protection.
  \item A \textbf{unified data correlation framework} that enables complex multi-source big data analysis and complex release gating criteria to prevent deployment-induced false positives.
  \item An \textbf{immutable multi-version deployment architecture} that allows continuous validation through shadow testing while maintaining production stability through deterministic rollback capabilities.
\end{itemize}
Unlike static detection systems that degrade over time as attackers develop new techniques, our architecture creates a self-reinforcing improvement cycle where each threat interaction contributes to stronger future defenses.
\begin{figure}[htbp]
\floatconts
  {fig:1}
  {\caption{Rapid response system architecture showing the end-to-end flow from threat intelligence ingestion to production deployment. Raw intelligence feeds the Threat Intelligence Platform, which generates detection signatures and training and validation data in the Data Platform, culminating in safe deployment through the Release Platform.}}
  {\includegraphics[width=0.3\linewidth]{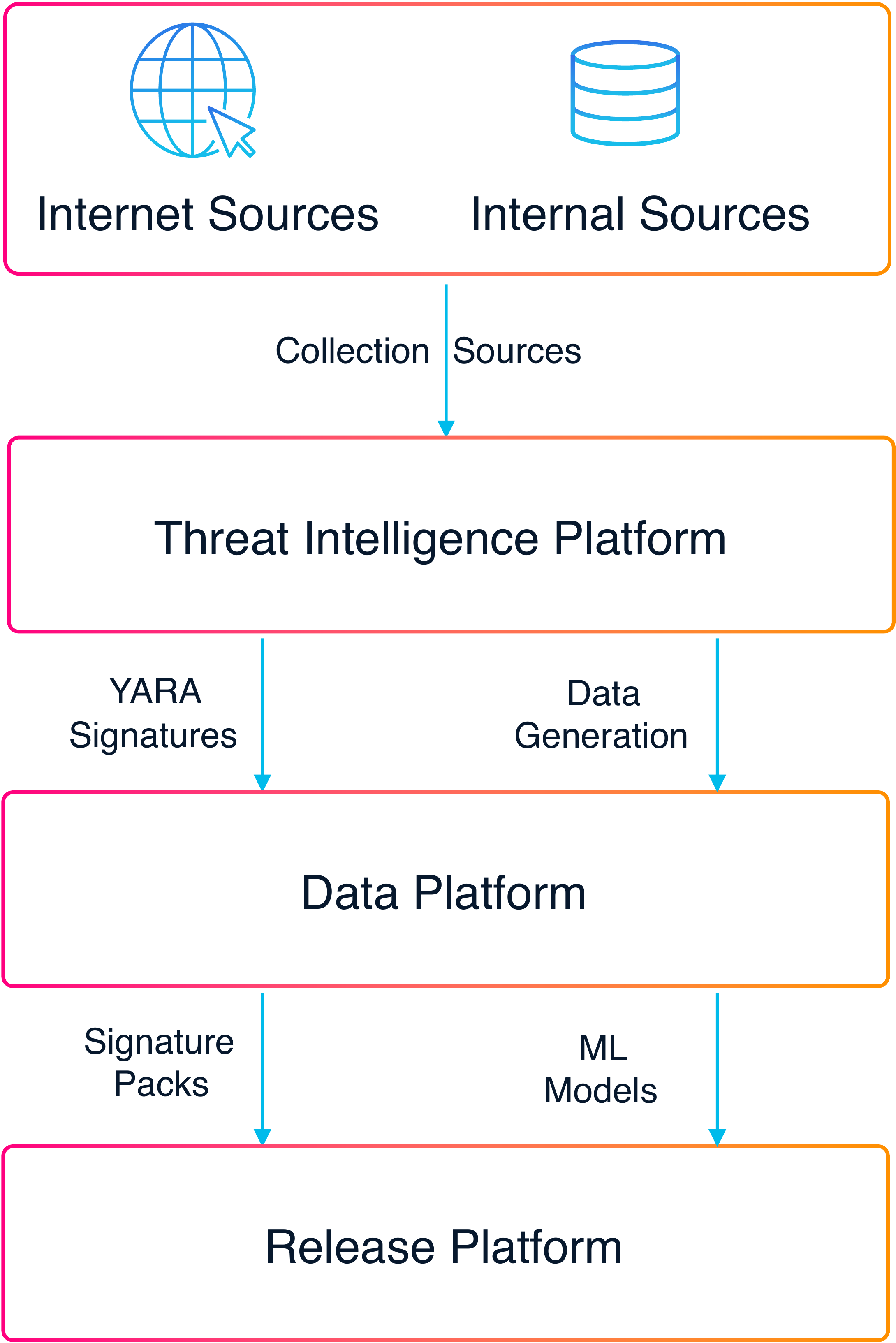}}
\end{figure}

\section{Threat Intelligence Operations}
Our threat intelligence operations capability serves as the first line of defense in our rapid response system, continuously monitoring the internet for LLM threats.
This system transforms raw intelligence on attacks and vulnerabilities into actionable protections through a pipeline that emphasizes automation, prioritization, and rapid signature deployment.
Given that it is not possible to predict which attacks will appear in the wild, we leverage a prioritization methodology to focus our efforts on attacks that are more likely to be observed.
The platform prioritizes threats based on implementation feasibility, attack practicality, and similarity to known attacks, while providing actionable outputs in the form of detection signatures and data generation modules.

We refer to this operational capability and its supporting tools as IntelOps.

\subsection{Architecture Overview}

The Threat Intelligence Platform consists of five primary components that operate in a continuous cycle:

\begin{enumerate}
    \item \textbf{Automated Collection and Monitoring}: Continuous ingestion of threat data from multiple open-source intelligence (OSINT) and closed sources.
    \item \textbf{Prioritization and Analysis}: Scoring and classification of threats into an analyst queue.
    \item \textbf{Analysis and Reporting}: Automated initial triage and reporting with human review.
    \item \textbf{Detection and Data Generation}: Development of detection signatures (e.g., YARA rules, which can detect patterns based on textual characteristics) for immediate defense and attack implementations for ML model training.
    \item \textbf{Feedback:} Routine review of detection hits and misses in Data Platform telemetry.
\end{enumerate}

Central to our Threat Intelligence Platform is a comprehensive and dynamically updated taxonomy~\cite{cisco2025taxonomy} that unifies existing AI security frameworks while addressing the unique operational needs of production LLM security.
We developed this taxonomy through a systematic review of established security standards, including the OWASP LLM Top 10~\cite{owasp2025}, MITRE ATLAS~\cite{mitreatlas}, and National Institute of Standards and Technology (NIST)'s Adversarial Machine Learning Taxonomy~\cite{nist2025}.
Unlike existing taxonomies that focus primarily on security or safety concerns, our framework bridges both domains, recognizing that production systems must defend against both malicious attacks and harmful outputs.
Each threat category includes detailed subcategories with explicit mappings to both OWASP and MITRE classifications, ensuring compatibility with existing security workflows while providing the granularity needed for detection purposes.
This granularity is maintained through an agile update mechanism that incorporates insights from emerging threats and incident analysis, allowing us to quickly integrate new edge cases and novel attack techniques.
This taxonomy serves as the basis for our Priority Intelligence Requirements (PIRs), which form a prioritized matrix of variables that guide our intelligence collection efforts~\cite{PIRs}.

The PIRs are structured around two key characteristics:

\begin{itemize}
    \item \textbf{Model- or application-specific vulnerabilities}: Threats targeting particular LLM architectures or implementations;
    \item \textbf{Tactics, Techniques, and Procedures (TTPs)}: Attack patterns and methodologies employed against LLMs.
\end{itemize}

Each PIR is assigned a priority score, correlating with its likelihood as a target (for example, a frontier LLM is more likely to be targeted due to a higher potential payload of successful compromise, thus a higher priority) or likelihood and potential impact of a TTP (for example, indirect prompt injection is high priority due to its frequency of use in the real world, and its potential impact if successfully deployed against a target).
Given the ever-evolving threat landscape around AI security, these priorities are reviewed and updated regularly to ensure alignment with emerging threats and customer needs.
Time-bound PIRs can also be added for temporary tracking and prioritization (e.g., 30-, 60-, or 90-day) windows for a particular threat or specialized datasets.

\begin{figure}[htbp]
\floatconts
  {fig:2}
  {\caption{Threat Intelligence Platform architecture showing automated collection from multiple sources (OSINT, academic research, internal findings), followed by prioritization scoring, human analyst review, and conversion to actionable protections through signature development and attack dataset generation.}}
  {\includegraphics[width=0.8\linewidth]{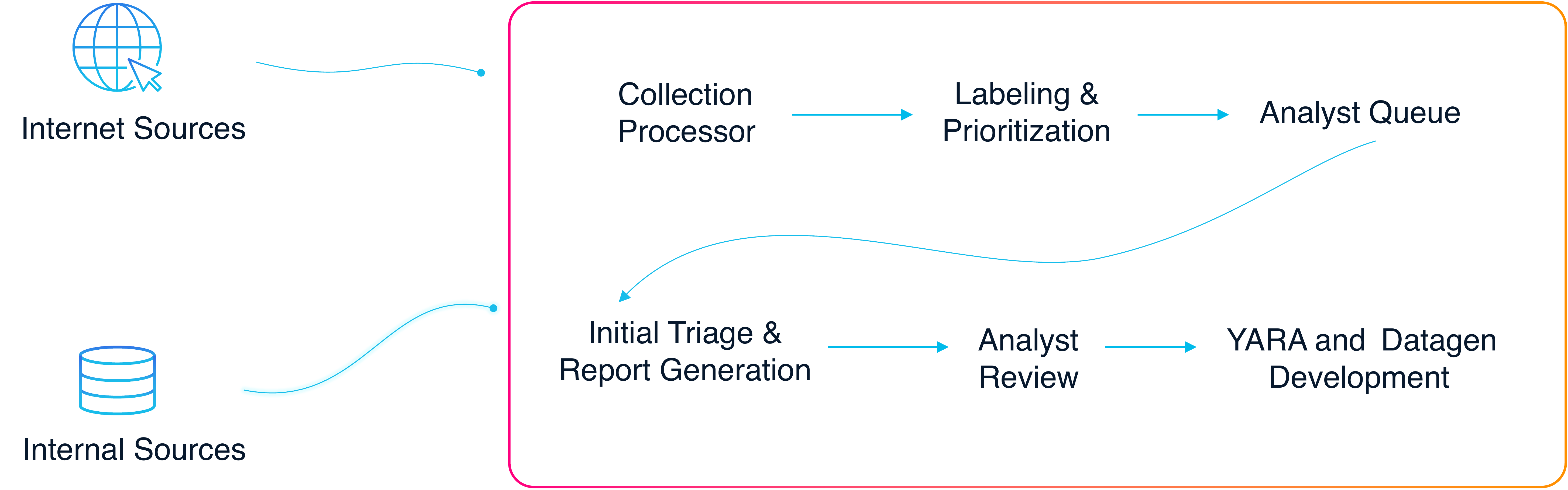}}
\end{figure}

\subsection{Automated Collection Pipeline}

The collection infrastructure conducts a nightly automated process that aggregates threat intelligence from multiple sources over the past 24 hours and includes (but is not limited to):

\begin{itemize}
    \item \textbf{Academic Research}: arXiv papers in relevant categories (cs.AI, cs.CR, cs.CL);
    \item \textbf{Security Feeds}: security research blogs, vulnerability databases, and threat intelligence vendors.
\end{itemize}

In addition to automated collections, raw intelligence can also come from:
\begin{itemize}
    \item \textbf{Ad hoc ingestion:} Individual sources and reporting added by a human analyst;
    \item \textbf{Internal Research:} Novel research performed by internal company teams.
\end{itemize}

The system employs deduplication mechanisms to prevent redundant processing, and we extract and summarize the text with Jina Reader~\cite{jina_reader}, ensuring consistent and complete content retrieval.

\subsection{Filtering and Scoring}

Upon collection, each source undergoes automated analysis against our PIRs.
Sources scoring above our defined thresholds are automatically flagged for human analyst review, while lower-scoring items are archived for potential future reference, allowing analysts to concentrate efforts on highest priority concerns first.
All LLM-generated artifacts, whether a label or a full report, can be manually edited or corrected by a human analyst via a web application, ensuring human-in-the-loop quality checks.

As intelligence comes in, our scoring algorithm quantifies the risk or severity of a potential threat or vulnerability:

\begin{equation}
P = \frac{T_{\mathrm{avg}} + 0.5\,M_{\mathrm{avg}} + S + 0.5\,E}{3}
\label{eq:pir}
\end{equation}

where:
\begin{itemize}
    \item $T_{\mathrm{avg}}$: Average priority scores of affected models and TTPs; if a model or TTP is not on the prioritized list, default to the lowest score.
    \item $M_{\mathrm{avg}}$: Average priority score of affected models and/or applications.
    \item $S$: Source credibility assesses the reputation of the source of intelligence (i.e., where the vulnerability or threat was discovered), from lowest to highest.
          A lower score means that it is less credible (e.g., arXiv has a lower score because the publications are not peer-reviewed and sometimes theoretical).
    \item $E$: Ease of implementation factors, calculated as the combination of \textbf{susceptibility} to an attack, whether signature-based detection opportunities are present, and whether datasets or data generation code is available for reference or for model training.
\end{itemize}

\textbf{Susceptibility} scores capture both a model's vulnerability to attack and the likelihood of exploitation. The scores range from Unlikely to Use/Difficult to Exploit to Highly Likely to Use/Trivial to Exploit, with intermediate values representing varying degrees of risk and implementation difficulty. This measure incorporates both technical feasibility and threat actor motivation. For example, if an attack simply requires a prompt template, the technical feasibility would be high, whereas an attack requiring fine-tuning a helpful attacker model and/or access to a large amount of compute may be a barrier to entry for less skilled or well-resourced threat actors.

The algorithm weighs TTPs ($T$) as the primary factor (full weight, coefficient of 1), while models ($M$) and ease of implementation ($E$) receive half weight (coefficient of 0.5), reflecting that the specific attack techniques are generally more indicative of immediate threat relevance than the targets or implementation difficulty alone.
The final score is normalized by dividing by 3, resulting in a score range of 0 to 5.
The resulting PIR score is then used to prioritize remediation efforts, allocate resources, and inform decision-making regarding the identified threat.

\subsection{Threat Analysis and Report Generation}

Prioritized reports are added to an analyst queue.
When an analyst is ready to review a particular report, they can utilize an LLM-assisted initial triage and report generation feature which also creates any relevant tasks in an integrated ticketing system.

The LLM reads the previously extracted source text and generates a comprehensive initial report that includes the following elements:

\begin{itemize}
    \item \textbf{Threat summary}: High-level summary with key takeaways of the vulnerability or attack technique;
    \item \textbf{Technical details}: Detailed explanation of the implementation mechanics;
    \item \textbf{Potential impact}: Assessment of risk to production LLM deployments, including affected models and reported Attack Success Rates (ASR) of that particular attack technique;
    \item \textbf{Example of attack:} If relevant, step-by-step instructions on implementing the attack and what the resulting attack prompt looks like;
    \item \textbf{Ease of implementation:} Details on what skills and resources are needed;
    \item \textbf{Detection and mitigation measures}: Initial suggestions for detection strategies and mitigation measures, including whether or not a detection signature (such as YARA) is appropriate.
\end{itemize}

Despite careful prompting for the automated initial triage component, human expertise and judgment remain crucial.
Security analysts review and improve on the automated reports by:

\begin{itemize}
    \item Validating the technical accuracy and completeness of initial reporting against the original source;
    \item Refining signatures for optimal detection coverage;
    \item Identifying edge cases and potential false positive scenarios;
    \item Assessing operational impact on customer workflows;
    \item Comparing attacks/vulnerabilities to any previously analyzed reports where existing detection coverage may be sufficient.
\end{itemize}

The Threat Intelligence Platform automatically publishes the generated reports to our internal ``IntelHub'' knowledge base for human review and finalization, creating a searchable repository of threat intelligence accessible to security, engineering, and product teams.

\begin{figure}[htbp]
\floatconts
  {fig:3}
  {\caption{Screenshot of IntelOps queue front-end that includes date of ingestion, title of source, affected models, TTPs, attack success rates, and analyst triage status, with additional filtering capabilities}}
  {\includegraphics[width=1\linewidth]{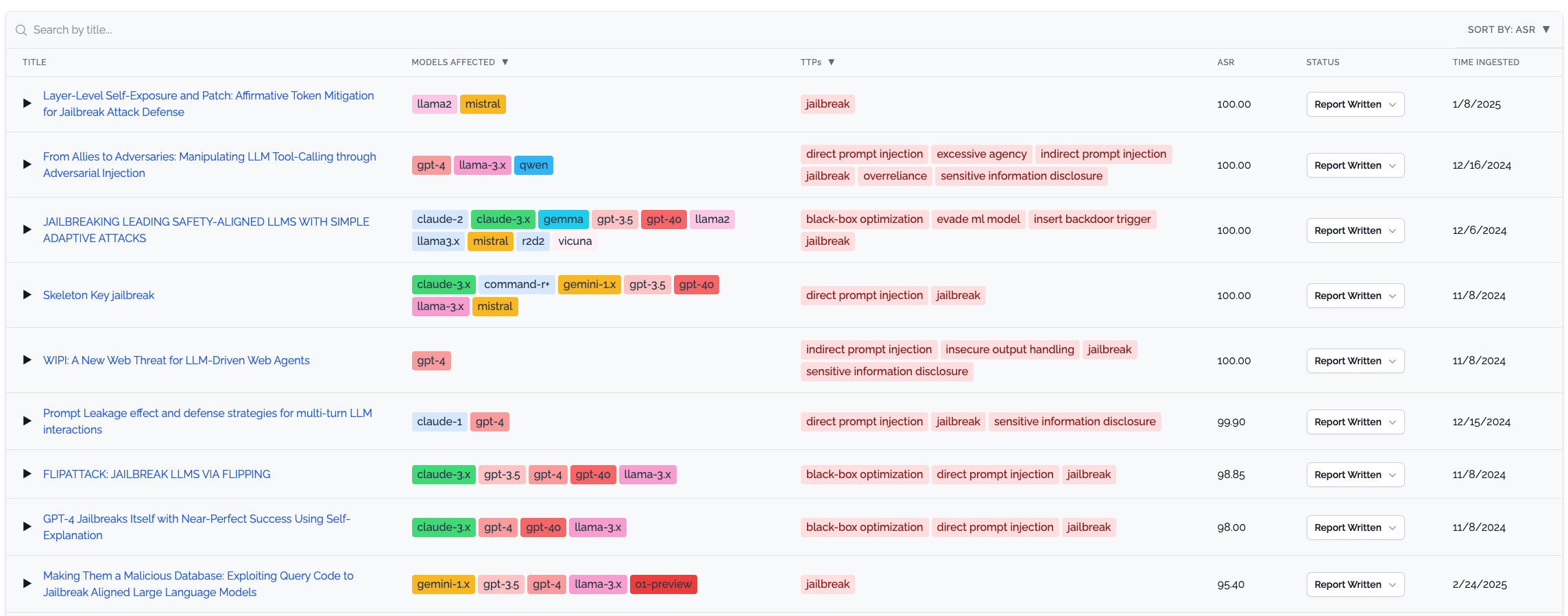}}
\end{figure}

\subsection{Rapid Signature Development}

One of the key components of our IntelOps process is the emphasis on rapid signature deployment as an immediate mitigation strategy.
While ML models provide superior generalization to protect against AI attacks, they require extensive retraining cycles that can be both time and resource intensive.
YARA, a tool originally designed to help identify and classify malware samples, can be utilized to provide immediate protection against newly discovered threats without having to retrain an ML model.
YARA rules also lend themselves well to certain types of prompt attacks (e.g., template-based, specific attack patterns) as they can combine textual, hexadecimal, and regular expression patterns with boolean conditional matching.

The signature development process is as follows:

\begin{enumerate}
    \item \textbf{Pattern identification}: Automated identification of unique attack patterns that could be well-suited for YARA
    \item \textbf{Signature generation}: Human development of YARA rules that capture the essential characteristics of the attack with enough granularity as to minimize false positives
    \item \textbf{Validation testing}: Signatures are tested against:
    \begin{enumerate}
        \item Our Data Platform's prompt corpus to ensure they do not inadvertently block legitimate use cases; and
        \item Internal signature metadata and formatting requirements.
    \end{enumerate}
    \item \textbf{Deployment readiness}: Validated signatures are packaged for immediate deployment through our release system and separately uploaded to the Data Platform for further use.
\end{enumerate}

See more information about safe signature deployment strategy in Section~\ref{sec:signature_updates}.

\subsection{Attack Data Generation}

While YARA signatures provide immediate protection against newly discovered threats, comprehensive defense requires training ML models on diverse attack data.
Our Threat Intelligence Platform employs a Python-based automated attack data generation framework that systematically transforms theoretical vulnerabilities into practical training datasets.
While this paper only discusses data generation for defensive model training purposes at a high-level, the same framework is also used to support our team's LLM red teaming and research projects.

The data generation system operates on a simple principle: process input datasets containing harmful intents and apply various attack techniques to produce adversarial prompts and conversations.
This approach enables scaling of training data while ensuring coverage of emerging attack patterns identified through our threat intelligence collection, and significantly reduces the time between threat identification and ML model updates.
Rather than waiting to collect sufficient in-the-wild attack samples, we can proactively generate training data for newly discovered techniques while maintaining the authenticity needed for robust model training.

Key capabilities include:
\begin{itemize}
    \item \textbf{Technique application}: Automated application of attack techniques such as jailbreaks, prompt injections, obfuscation, and multi-turn strategies to base intents;
    \item \textbf{Multi-turn generation}: Creation of conversational attacks that attempt to achieve malicious goals through seemingly benign dialogue progression;
    \item \textbf{Parallel processing}: Multi-worker architecture with checkpointing to handle large-scale dataset generation efficiently;
    \item \textbf{Metadata preservation}: Comprehensive metadata including technique used, harm category, and generation parameters for each generated attack.
\end{itemize}

When analysts identify new attack patterns through the threat intelligence pipeline, these techniques are quickly integrated into the generation framework as plugins.
This ensures that our ML models receive training data that reflect the latest threat landscape.

\subsection{Human Labeling and Feedback}

Finally, human expertise remains essential in our process, serving as both a quality control mechanism and a source of ground truth for our automated systems.
While automated labeling improves the scalability of this capability, human labeling ensures accuracy, catches edge cases, and identifies systematic issues that might otherwise go unnoticed in purely automated pipelines.
Human labeling serves as a critical checkpoint for monitoring the performance of our automated labeling algorithms.
By maintaining a continuous human review process, we can:
\begin{itemize}
    \item \textbf{Detect label drift}: Identify when automated labeling begins to diverge from human consensus, indicating potential model degradation or shifts in attack patterns
    \item \textbf{Diagnose false positives}: Determine whether detection errors stem from model generalization failures or misaligned consensus labeling criteria
    \item \textbf{Calibrate consensus algorithms}: Adjust automated labeling instructions based on patterns identified by human review
    \item \textbf{Refine our threat taxonomy}: Human analysts regularly encounter edge cases that challenge existing categorizations, providing insights that sharpen distinctions between similar attack categories, identify emerging attack patterns that do not fit existing classifications, or reveal common misunderstandings that customers may also experience
\end{itemize}

This intelligence platform establishes a dynamic and adaptive framework for understanding, anticipating, and mitigating threats against AI models and systems.
Its iterative nature and emphasis on both automation and human expertise are critical for maintaining a resilient defense against novel threats and attack techniques.

\section{Data Platform}

The goal of the Data Platform is to provide a single location for all data storage, aggregation, enrichment, labeling, and decision making.
Information from multiple sources is systematically aggregated and correlated, ensuring comprehensive artifact analysis through consolidated data representation.
This architectural approach enables cross-source information utilization for enhanced decision-making processes, including detection model evaluation, resource prioritization, and automated enrichment workflows.

The Data Platform leverages the distributed data warehouse Snowflake, which prioritizes adaptability over traditional processing paradigms, as shown in Fig.~\ref{fig:platform}.
A key architectural decision was choosing between traditional ETL pipelines and our novel warehouse-centric approach.
Rather than implementing conventional ETL pipelines with rigid schemas and predetermined processing sequences, which can create brittleness when adapting to new threat types, our design emphasizes flexible data transformation capabilities that can rapidly adapt to evolving requirements.
This approach supports dynamic threat response through accelerated operational cycles while maintaining system stability.

Once the initial raw data is ingested, core processing occurs within the warehouse environment using standard query languages that are extended via user-defined functions (UDFs), or UDFs that call out to external services.
This enables augmentation of native capabilities for any functionality that is not directly available in SQL (e.g., calling out AI services or applying customer Python code).

The unified data storage model enables logical pipeline definitions that eliminate physical structure dependencies when aggregation or data collection methods evolve.
Unlike traditional approaches that involve explicit pre-ingestion processing or external data augmentation, our design allows for rapid adjustments.
With all primitives defined in SQL, we can quickly mix and match them to modify the pipeline without having to adjust or redeploy any additional infrastructure.

\begin{figure*}[ht]
\floatconts
  {fig:platform}
  {\caption{Data Platform}}
  {%
    \subfigure[Data pipeline architecture for LLM prompt processing, showing iterative enrichment workflow from raw data ingestion through correlation and labeling resulting in a unified table, enabling flexible business logic implementation through SQL views.]{\label{fig:top-part}%
      \includegraphics[width=0.8\linewidth]{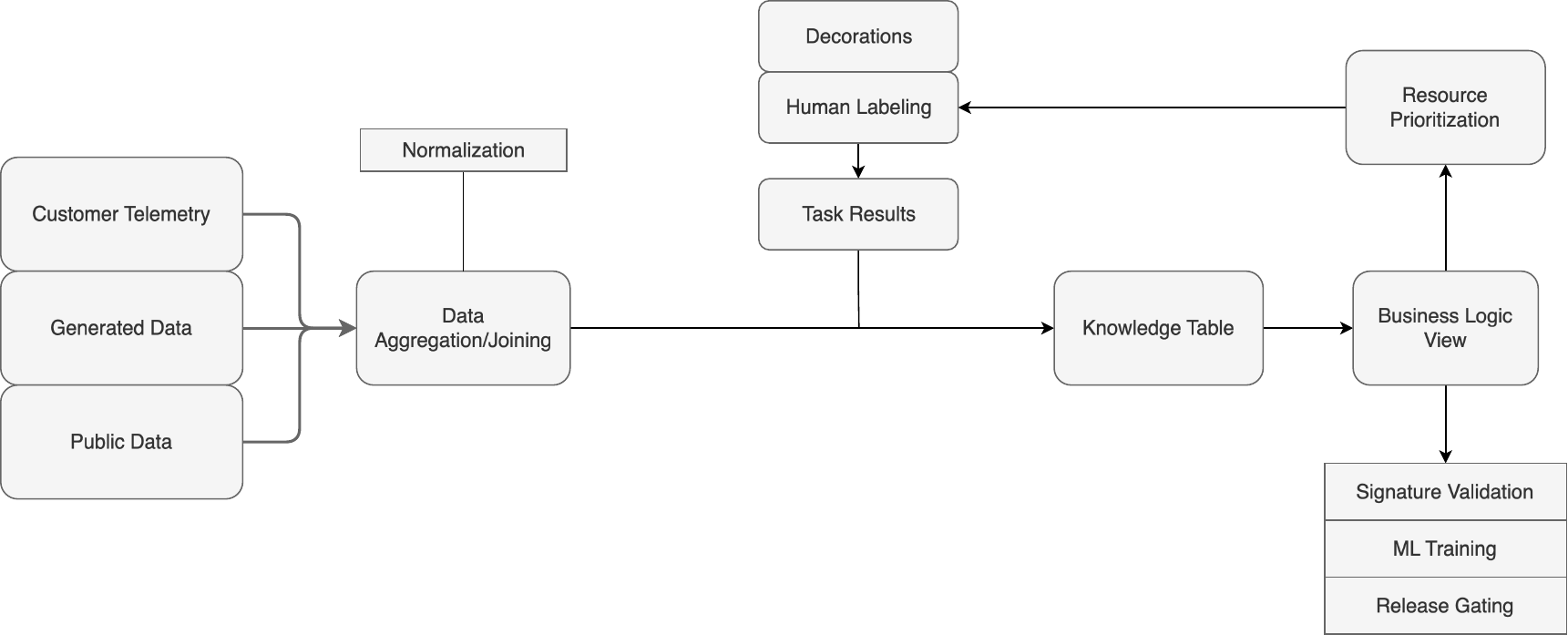}}

    \subfigure[Snowflake-based Data Platform architecture illustrating SQL-centric processing augmented with Python UDFs and external service calls, providing extensible primitives for flexible data pipeline operations and cross-source correlation.]{\label{fig:bottom-part}%
      \includegraphics[width=0.8\linewidth]{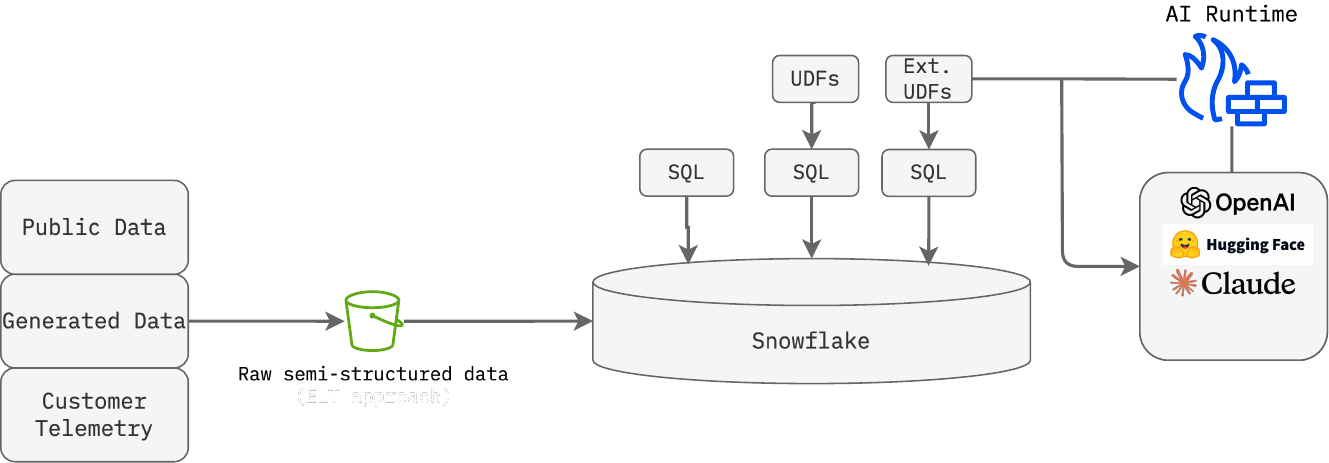}}
  }
\end{figure*}

Our Data Platform regularly aggregates and correlates the following resources:
\begin{itemize}
    \item Customer telemetry, including the guardrail response and the customer prompt (when allowed);
    \item All known publicly available datasets on the internet that are deemed relevant (e.g., open-source datasets in Hugging Face and GitHub);
    \item Internally generated human labels;
    \item Model-generated labels;
    \item Prompt translations into various languages;
    \item Internally generated data, such as data produced by the threat intelligence pipeline described in the previous section; and
    \item Detailed responses from current and upcoming versions of the guardrails, including ML scores and signature matches.
\end{itemize}

Customer telemetry is processed according to customer data handling policies; we anonymize and minimize PII where applicable.

Beyond standard data aggregation and correlation, our Data Platform handles two critical tasks.
First, it identifies any additional data needed, prioritizing outstanding computation tasks in a queue, and then collects the required information as a background task.
Second, it computes the golden labels for a given artifact based on all the available information.

\subsection{Prioritization Tasks}

The prioritization framework operates through a two-phase approach: first, defining the prioritization queue, and batch processing execution.
The queue uses dynamic data integration that combines processed knowledge with current operational data through database views.
Selected workloads are then processed in batches with results systematically recorded through standard database operations.

Our pipeline runs prioritization tasks on all known prompts in our database, which include:
\begin{itemize}
    \item Multi-language processing capabilities for enhanced detection coverage;
    \item Automated labeling workflows for training data quality improvement;
    \item Language classification modules for content categorization and processing optimization;
    \item Performance evaluation against current and upcoming guardrails for offline performance evaluation and release gating;
    \item Scoring prompts against third-party guardrails for performance monitoring and labeling improvements; and
    \item Scanning all prompts against our signatures, used for monitoring and identifying issues or gaps.
\end{itemize}

These processes continue until all data is computed and the prioritization updates automatically as new data propagates into the knowledge table.
Importantly, with all data available for each prompt in the knowledge table, we can make complicated and rapidly changing prioritization decisions despite information being aggregated across multiple sources.

The knowledge table is a unified view keyed by prompt ID that aggregates labels, model scores, signature matches, metadata, and provenance for decision-making.

\subsection{Labeling}

Labeling is the most important output produced by our data pipeline, as it is the most impactful lever that one can leverage to increase ML detection accuracy, to validate that upcoming releases are safe to deploy, and to quickly identify detection issues in production.
However, labeling faces two major challenges.
First, benign data labeling must be extremely accurate.
Any deployed detection system requires a very low false positive rate; having a labeling error higher than the desired false positive rate would be unworkable.
Second, we must be able to identify attacks, especially novel ones.
Relying on a single labeling tool would limit our detection system's value, offering no improvement beyond that tool's inherent capabilities.

Given the volume of data flowing through our platform, strategic prioritization that highlights human labeling efforts is essential.
Our Data Platform automatically prioritizes samples for human review, such as detections from actual customer telemetry or samples with low or borderline confidence scores.
We sample detections and near-misses for review proportional to their risk, novelty, and potential customer impact.
Effective labeling demands both a wide breadth of data to label and an acceptance of its inherently iterative nature.
Our infrastructure achieves these two conditions by aggregating information from multiple instructed LLMs, human labeling efforts, signature detections, and original source labels.

\section{Release Platform}

One of the largest challenges in deploying a detection and blocking system like our guardrails is updating its detection components post-release.
The primary risk lies in unforeseen shifts in the detection distribution, which could potentially disrupt established customer workflows.
We have seen countless examples of security vendors accidentally generating catastrophic levels of false positives that ultimately break critical customer infrastructure~\cite{crowdstrike2024technical, sophos2012selfdetection, mcafee2010falsepositive}.
Our Data Platform is designed to minimize the chance of such an event, relying on three major components that need updating.
First are signatures, which we can release rapidly to remediate any ongoing issues or concerns.
Second are ML detection models; these take longer to update and require additional validation due to their potential to cause unexpected shifts in the detection distribution.
Third, the orchestration logic that integrates the previous components into the final detection (Fig.~\ref{fig:production}).

\begin{figure}[htbp]
\floatconts
  {fig:production}
  {\caption{Release Platform architecture demonstrating immutable component deployment with concurrent versioning. The central orchestrator routes customer requests to appropriate guardrail versions, enabling seamless shadow deployments, gradual rollouts, and instant rollbacks while guaranteeing that already deployed guardrails cannot be accidentally disrupted during updates.}}
  {\includegraphics[width=1\linewidth]{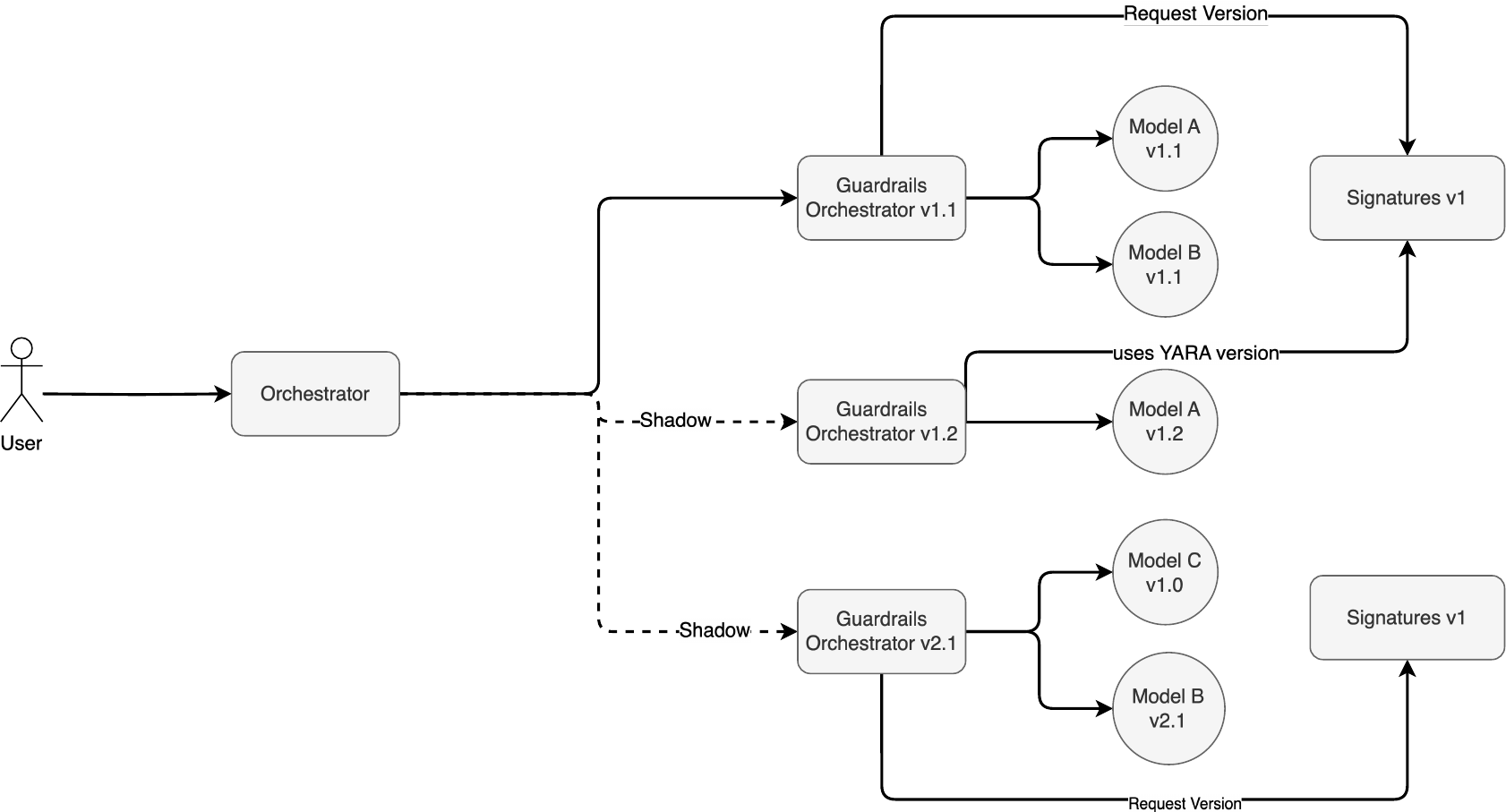}}
\end{figure}

The platform architecture supports simultaneous deployment of multiple versions of guardrails within the same deployment.
Each detection implementation and its associated components are entirely immutable, ensuring that no updates can disrupt existing detection behavior.
While there are no restrictions on the number of guardrail versions that can be deployed concurrently, it is typical to maintain both production and shadow deployments at all times.

Instead of updating existing guardrails, a new version is released alongside the previous one, rather than replacing them directly.
This approach enables gradual customer transition and simplified rollback procedures without the complexities of a conventional release cycle.
The deterministic relationship between inputs and outputs across components aids in reproducible results and enables efficient caching strategies for optimized performance.

This design enables updates to be pushed directly to production because they cannot impact existing production systems.
We thus unblock the engineering release process, enabling hot patches and other engineering updates to be released into production without being blocked by the lengthy, expensive, and extensive detection QA release processes described above.
The ML team can safely and thoroughly check for detection regressions across multiple version releases without interruption, even though the engineering version is being updated in the preview, staging, and production environments constantly.
Below, we detail how this architecture is used to safely update our detection.

\subsection{Signature Updates}
\label{sec:signature_updates}

Signatures, in this case expressed as YARA rules, are developed through the intelligence processes detailed earlier and serve as rapid response tools for critical detection issues that require immediate resolution or provide interim protection against emerging threats while ML models undergo retraining and redeployment cycles.
Our goal is a swift release of updates, starting from signature creation to deployment to an initial set of customers.

These signatures function as a first line of defense, remediating emerging threats or resolving impactful false positives and false negatives.
While not designed for the broad generalization capabilities of machine learning approaches, they deliver critical remediation during the ML model update cycle.
The primary advantage of rule-based detection lies in its predictable impact on overall detection distributions, allowing for more confident deployment compared to probabilistic ML detection approaches.

This rule-based approach provides deterministic behavior that complements machine learning detection capabilities, offering rapid response mechanisms for time-sensitive security updates while maintaining system reliability and customer confidence.
We validate new signatures to ensure they:
    \begin{itemize}
        \item Accurately identify expected malicious patterns with precision;
        \item Minimize false positive rates on legitimate use cases;
        \item Comply with internal standards for metadata and formatting requirements; and
        \item Operate within established performance and resource utilization thresholds.
    \end{itemize}

The signature release process is designed to balance speed with safety through a structured multi-environment deployment pipeline.
The release workflow progresses through three distinct environments:
\begin{enumerate}
    \item \textbf{Internal Snowflake Environment}: An initial PR request.
          The updated signature is scanned against all prompts in the data platform and an automated report is generated.
          The deltas between the previous signature package and the new package are inspected to ensure no unexpected outcomes are observed.
    \item \textbf{Preview/Staging Environment}: Initial deployment for functional validation.
          A large prompt corpus is re-scanned with the full guardrails, which now include the new signatures.
          Since the individual detection components making up the guardrails are immutable, the ML scores are pulled from cache rather than recomputed, thus the process is fairly fast and low cost.
    \item \textbf{Production Environment}: Gradual roll-out to customer deployments.
          The process can be further augmented with an optional shadow deployment if additional checks are required.
\end{enumerate}

Each environment transition requires specific validation criteria to be met, ensuring that signature updates maintain detection quality while avoiding disruption to legitimate use cases.
Signature updates are managed through a version control system where each release is tagged with a unique identifier, and multiple previous versions are also available in all environments.
This approach provides a complete audit trail of all signature changes, instant reversion to previous versions if issues arise, and enables multiple signature versions to coexist during transition periods.

Post-deployment monitoring through our data platform provides feedback on signature effectiveness, including detection rates and false positive metrics that are tracked in real-time, as well as customer impact assessment through telemetry analysis.
Post-deployment reviews allow us to further refine signatures and ML model training priorities.

\subsection{Model and Logic Updates}

Updating ML detection models and associated orchestration logic represents a higher-risk operation due to the potential for unexpected detection behavior changes that could disrupt established customer workflows.
Risk mitigation involves implementing a multi-stage release process that extends over several days to weeks, enabling validation at each phase.

Our staged process, described below, ensures detection quality while minimizing operational disruption, providing multiple validation checkpoints and maintaining system reliability throughout the update process:

\begin{enumerate}
\item New models and guardrail logic are committed to master, creating a new version of guardrails.
\item The new guardrail version is deployed into pre-production environments alongside the current version of guardrails.
\item The new guardrails are evaluated against a curated and prioritized prompt dataset in the Data Platform.
\item Results undergo review by the threat intelligence team and ML owners, focusing on differential analysis between versions.
\item Signatures are updated to address systematic false positives or other classification errors identified during evaluation.
\item Following issue resolution to acceptable thresholds, the new guardrails are rolled into production in shadow mode.
\item Comprehensive performance metrics are analyzed in the Data Platform, examining detection patterns and any increases in flag rate on synthetic or previously captured prompts when actual customer prompts are not available.
\item Customers are then gradually migrated to the new guardrail version, maintaining rollback capabilities for rapid issue response.
\end{enumerate}

Release gating evaluates changes in false positive rate at the operating point, recall, and flag rate deltas, with shadow performance informing promotion decisions.

\section{Previous Work}

As discussed earlier in this paper, significant research has focused on developing detection methods for LLM attacks and protections against them, consisting of technical controls and operational best practices~\cite{owasp2025, jiang2023identifyingmitigatingvulnerabilitiesllmintegrated, Christiano2022, bullwinkel2025lessonsredteaming100, jain2023baselinedefensesadversarialattacks}.
But relatively few works address the operational challenges of deploying these defenses in production environments requiring rapid threat response and continuous adaptation.
We have not found similar examples of a dedicated threat intelligence function designed specifically for AI-related threats that considers the expanded attack surface and the deepening integrations of generative AI and agentic AI tools in software systems and technology.

Early work in operationalizing LLM security began with the development of Vigil~\cite{vigil2023} in 2022, which introduced the application of YARA rules to the domain of LLM threat detection.
The open-source project demonstrated that pattern-matching techniques could effectively identify prompt injection attempts and other LLM-specific threats while leveraging existing security tooling rather than requiring specialized development of novel tools.
Our comprehensive defense platform described in this paper significantly extends this foundational approach by integrating rule-based detection into an enterprise-grade threat intelligence pipeline with automated deployment capabilities, enabling organization-scale protection with the ability to rapidly respond to emerging threats.

Recent advances in prompt injection detection have produced promising approaches, including research that compares the performance of early detection systems for LLM security~\cite{gakh2025enhancingsecurityllmapplications}.
Google DeepMind's CaMeL~\cite{debenedetti2025defeatingpromptinjectionsdesign} takes an architectural approach, applying traditional software security principles like control flow integrity to LLM systems.
PromptShield~\cite{jacob2025promptshielddeployabledetectionprompt} achieves detection with low false positive rates through fine-tuning detectors on both conversational and application-structured data, while Attention Tracker~\cite{hung2024attentiontrackerdetectingprompt} introduces training-free detection by analyzing attention patterns within LLMs. However, Hackett et al.~\cite{hackett2025bypassingpromptinjectionjailbreak} demonstrated that many prominent guardrail systems can be bypassed with up to 100\% success rates using character injection techniques, highlighting the limitations of point-in-time detection approaches.
While comprehensive prevention remains an open problem, our platform reduces the window of vulnerability by translating new threats into protections quickly through rapid signature updates and safe staged deployments.

\section{Conclusion}

This paper presents a comprehensive multi-layered defense strategy for protecting LLMs against evolving AI threats through rapid response mechanisms.
By synthesizing established cybersecurity approaches with novel AI-specific protection strategies, we have created a production-grade system that effectively balances the competing demands of detection speed, accuracy, and operational stability while providing resilient protection against both documented and novel threats.

Our integrated architecture, comprised of three interdependent platforms, addresses the complete threat response lifecycle:

\begin{enumerate}
    \item The \textbf{threat intelligence platform} provides continuous threat landscape monitoring and rapid protection through automated intelligence collection, threat prioritization, and detection signature development.
    \item The \textbf{data platform} enables intelligent decision-making through comprehensive data aggregation, intelligence labeling workflows, and prioritization algorithms.
    \item The \textbf{release platform} ensures safe deployment through immutable version management, progressive rollout mechanisms, and seamless rollback strategies.
\end{enumerate}

Production deployment across enterprise environments has validated the platform's effectiveness in protecting AI deployments while preserving the agility required to respond to new attack techniques and methodologies.
The system successfully bridges the gap between immediate tactical response and strategic model improvement initiatives, creating a unified defensive capability that addresses both reactive and proactive AI threats.

While our platform demonstrates significant advancement in LLM security capabilities, several areas present additional opportunities for continued improvement and research investment.
Future technical development could, for example, improve cross-model generalization to address the growing diversity of LLM architectures.
Threat intelligence processing capacity must also evolve to handle increasingly sophisticated attack patterns without overwhelming threat intelligence and security operations teams.
The development of more sophisticated algorithms and threat classification capabilities will be necessary to adapt to novel attack methodologies.

As LLMs become increasingly integrated into critical business processes and as agentic AI capabilities increase in scope, the need for robust security platforms will only grow.
LLMs, while transformative, represent new attack surfaces that, if compromised, could lead to significant data breaches, service disruptions, or reputational damage.
While no system can yet guarantee impenetrable protection against all conceivable adversarial inputs, our platform demonstrates that principled engineering combined with rapid response capabilities can help safely speed up the OODA loop, thus enabling us to provide enhanced protection.

The architectural principles and implementation strategies described in this paper are designed to serve as a foundational blueprint for organizations seeking to establish or enhance their own AI security capabilities.
By sharing our design decisions and operational insights, we hope to advance a collective understanding of protective measures for AI systems in production environments.
The future of AI security lies in adaptive systems that learn, evolve, and improve alongside the threats they defend against.
Our goal is to foster a collaborative environment where best practices can be universally adopted and continuously improved upon.


\bibliography{main}

\end{document}